\documentclass{article}
\usepackage[utf8]{inputenc}
\usepackage[margin=20truemm]{geometry}
\usepackage{cite}
\usepackage{amsmath,amssymb,amsfonts}
\usepackage{algorithmic}
\usepackage{graphicx}
\usepackage{textcomp}
\usepackage{amsopn,color,hyperref,cleveref,cases,algorithm,subfig,multirow,appendix}
\DeclareMathOperator{\diag}{diag}
\DeclareMathOperator{\argmax}{argmax}
\DeclareMathOperator{\argmin}{argmin}
\crefname{algorithm}{Algorithm}{Algorithms}
\Crefname{algorithm}{Algorithm}{Algorithms}
\crefname{equation}{Eq.}{Eq.}
\Crefname{equation}{Equation}{Equations}
\crefname{table}{Table}{Tables}
\Crefname{table}{Table}{Tables}
\crefname{figure}{Fig.}{Figures}
\Crefname{figure}{Figure}{Figures}
\Crefname{subsection}{subsection}{subsections}
\newcommand*{\reviewerA}[1]{\textcolor{black}{#1}}
\title{Greedy Sensor Selection for Weighted Linear Least Squares Estimation under Correlated Noise}

\author{
Keigo Yamada\thanks{the Department of Aerospace Engineering of Tohoku University, Sendai, Miyagi, Japan},\and
{Yuji Saito}\footnotemark[1],\and
{Taku Nonomura}\footnotemark[1],\and
{Keisuke Asai}\footnotemark[1],
}

\begin{document}
\maketitle
\begin{abstract}
Optimization of sensor selection has been studied to monitor complex and large-scale systems with data-driven linear reduced-order modeling.
An algorithm for greedy sensor selection is presented under the assumption of correlated noise in the sensor signals.
A noise model is given using truncated modes in reduced-order modeling, and sensor positions that are optimal for generalized least squares estimation are selected.
The determinant of the covariance matrix of the estimation error is minimized by efficient one-rank computations in both underdetermined and overdetermined problems. 
The present study also reveals that the objective function with correlated noise is neither submodular nor supermodular.
Several numerical experiments are conducted using randomly generated data and real-world data.
The results show the effectiveness of the selection algorithm in terms of accuracy in the estimation of the states of large-dimensional measurement data.
\end{abstract}



\section{Introduction}
\label{sec:introduction}
Observation is the primal step toward understanding real-world phenomena.
When monitoring quantities that cannot be observed directly, system representations are constructed for describing the dynamical behavior of the phenomena of interest as the state space of physical equations including unknown parameters. 
Therefore, parameter estimation using sensor measurements has long attracted attention in many engineering and scientific applications~\cite{wouwer2000approach, alonso2004optimal, kraft2013optimization, tzoumas2018selecting, krause2008efficient, yoshimura2020application}. 
A reduction in the number of measurements is concurrently demanded for more practical use, especially under resource constraints on sensors and communication energy, or for processing measurements in real time.
Optimization problems are presented here using metrics defined for sensor locations. The Fisher information matrix is a well-used metric for the assessment of uncertainty in parameter estimation, as this optimization task is closely related to the experimental designs~\cite{kincaid2002d, bates1996experimental, padula1998optimal}. 
The theory of information and statistics are also informative criteria for the optimization task~\cite{chepuri2015sparse, krause2008near, garnett2010bayesian, paris2021robust, semaan2017optimal}.

\reviewerA{Sensor placement based on the physical equations has been adopted in the reconstruction of physical fields, such as sound or seismic wave distribution~\cite{douganccay2009optimal,du2014optimal,taghizadeh2015robust}.
With a similar formulation of the placement, sensor selection has been conducted by choosing the best subset of sensor nodes in the context of network monitoring and target tracking~\cite{phatak2001recursive, aggarwal2017sensor}.}
Recently, advances in data-driven techniques enable us to obtain a system representation from astonishingly high dimensional monitoring data for complex phenomena, with sensor nodes defined by each sampling point in the data~\cite{berkooz1993proper, schmid2010dynamic, rowley2004model, brunton2019data, kutz2016dynamic, brunton2015compressed,inoue2022data}. 
Spatiotemporal correlations between measurements at sensor nodes are here represented as a superposition of a limited number of bases, sometimes with impositions of a physical structure or robustness to the system~\cite{baddoo2021physics,takeishi2017bayesian,scherl2020robust,brand2002incremental}.
The use of optimized measurement has been accelerating the applications of data-driven modeling in several engineering fields such as face recognition~\cite{manohar2018data}, inbetweening of non-time-resoloved data~\cite{tu2013integration},  noise reduction~\cite{inoue2021data}, state estimation for air flow~\cite{kanda2021feasibility}, wind orientation detection for vehicles~\cite{inoba2021}, and source localization~\cite{kaneko2021data}. 

The main challenge of such optimization problems is intractability, where the problems are often classified as being nondeterministic polynomial-time hard.
Therefore, heuristics to find suboptimal solutions have been intensely discussed. 
For example, the selection problem is solved by the linear convex relaxation methods~\cite{joshi2009sensor, nonomura2021randomized, chmielewski2002theory} or by using proximal gradient methods~\cite{dhingra2014admm, nagata2021data}.
The submodularity property in these optimization problems also encourages the use of greedy algorithms~\cite{nemhauser1978analysis, feige2011maximizing, golovin2011adaptive, hashemi2020randomized,saito2020data,saito2021determinant,saito2021data}.
Some recent studies attempt to improve the performance of greedy algorithms by grouping and multiobejective strategies, which considers multiple sensor subsets simultaneously~\cite{jiang2019group, nagata2021randomized,nakai2022nondominated}.

\reviewerA{Despite these established methodologies, considering the complex structure of measurement noise still remains a great challenge for the sensor selection problem, as treated in recent studies~\cite{pazman2001optimal,jamali2014sparsity,masazade2012sparsity,rigtorp2010sensor}.
The discrepancy between measurements and the model should be surely considered as spatially correlated measurement noise, for example, due to numerical computation of equations and assumptions in the modeling~\cite{kirlin1985optimal,yu2018scalable}, or due to truncation in the model order reduction~\cite{yamada2021fast,nagata2022data}. 
These errors cause correlated effects on the estimation adversely, thus the evaluation of the measurement noise should be included in the sensor selection objective.
%
As previously illustrated by Ucinski~\cite{ucinski2020d}, a sensor selection problem with continuous relaxation loses convexity when the noise covariance term is introduced.
Liu {\it et al.}~\cite{liu2016sensor} put forward a semidefinite programming for the sensor selection problem with spatially correlated noise, but the calculation becomes prohibitive due to the large problem size. 
The greedy algorithm introduced in this study smoothly integrates the noise covariance into the formulation in Saito {\it et al.}~\cite{saito2021determinant}, although the loss of submodularity is also confirmed.
Another advantage of greedy methods is to circumvent the rounding procedures for obtaining sensor positions from a relaxed solution, which are still an arguable process especially under the nonconvexity of the optimized objective function.}

The objective functions for the greedy selection are derived in both of the overdetermined and underdetermined settings, which generalize the previous D-optimality-based formulation~\cite{saito2021determinant}.
The Fisher information matrix is defined in this work to evaluate the uncertainty of linear least squares estimation for a static system.
The algorithm leverages one-rank computation for both a sensitivity term for each sensor and a weighting term for measurement noise.
Some of the recent studies introduced prior distribution for Bayesian estimation~\cite{liu2016sensor, yamada2021fast}, maximum a posteriori estimation~\cite{shamaiah2010greedy}, and Kalman filtering~\cite{zare2018optimala}.
Virtually, the hyper parameters in those distributions must be optimized using some information criteria or cross-validation techniques.
The formulation in the present study excludes a prior distribution, because the optimization for hyperparameters is difficult for high dimensional data treated in \cref{section: noaasst}.
In summary, we herein
    1)~propose an optimization problem for greedy sensor selection generalized for correlated measurement noise, which is easily extended to various optimality criteria, 
    2)~confirm that the objective function is neither submodular nor supermodular, and
    3)~formulate a fast greedy algorithm that selects sensors optimized for both underdetermined and overdetermined cases in the weighted linear least squares estimation.
\section{Formulation and Algorithm}
This section describes problem settings for sensor selection tailored for weighted least squares estimation.
Then, algorithms for greedy selection are discussed.
\subsection{Sparse Sensing} \label{section: sparse}
{A linear measurement equation for $p$ sensors $\mathbf{y} \in \mathbb{R}^{p}$ and a state vector of $r$ components $\mathbf{z} \in \mathbb{R}^{r}$ is corrupted by Gaussian noise $\mathbf{w}\sim\mathcal{N}(\mathbf{0}|\mathbf{R})\in \mathbb{R}^{p}$, which is independent of $\mathbf{z}$,
\begin{align} 
\mathbf{y}=\mathbf{C}\mathbf{z}+\mathbf{w}. \label{eq: measurement}
\end{align}
We assume that the sensor characteristic $\mathbf{C} \in \mathbb{R}^{p{\times}r}$ is known in advance and nonsingular, and the covariance of the measurement noise $\mathbf{R} \in \mathbb{R}^{p{\times}p}$ is positive definite and symmetric.
An parameter vector $\tilde{\mathbf{z}}$ is estimated from \cref{eq: measurement}:
\begin{subnumcases}{\tilde{\mathbf{z}}=}\label{eq: estimation gls}
    \mathbf{C}^{\top}\left(\mathbf{C}\mathbf{C}^{\top}\right)^{-1}\mathbf{y}\, & $\left( p\le r \right)$ \label{eq: estimate noise under}\\
    \left(\mathbf{C}^{\top}\mathbf{R}^{-1}\mathbf{C}\right)^{-1}\mathbf{C}^{\top}\mathbf{R}^{-1}\mathbf{y}\, & $\left( p > r \right)$\label{eq: estimate noise over}, 
\end{subnumcases}
Note that Eq.~\eqref{eq: estimate noise under} corresponds to the minimal norm solution in which the measurement noise is not considered, though Eq.~\eqref{eq: estimate noise under} is derived from the formulation including measurement noise. On the other hand, Eq.~\eqref{eq: estimate noise over} is a minimum variance unbiased estimation considering measurement noise as the generalized least squares estimation \cite[Section~4.5]{steven1993fundamentals}.
The present study focuses on the formulation above, excluding any prior distribution of the state variables.}

We also assume a large number of possible measurement points, {\it e.g.}, $\mathbf{x}\in\mathbb{R}^{n}\, (n\gg r)$.
The linear coefficients and noise covariance for all of the measurement points are expressed as $\mathbf{U}\in\mathbb{R}^{n\times r}$ and $\mathcal{R}\in\mathbb{R}^{n\times n}$, respectively.
Actual calculations for these terms are introduced later herein at \cref{section: modeling}.
With these notations, the measurement is expressed by substituting $\left(\mathbf{y},\,\mathbf{C},\,\mathbf{R}\right)\leftarrow\left(\mathbf{x},\,\mathbf{U},\,\mathcal{R}\right)$.
Here, the estimation \cref{eq: estimation gls} is redundant if $r$ is small, and thus the reduced measurement $p\ll n$ is sufficient in terms of both estimation quality and calculation efficiency.
{A sensor indication matrix $\mathbf{H}{(\mathcal{S}_{p})} \in \mathbb{R}^{p{\times}n}$ is defined for a set $\mathcal{S}_{p}$ of $p$ sensor indices selected from $n$ candidates.
The position of unity in the $i$-th row of $\mathbf{H}{(\mathcal{S}_{p})}$ is associated with the $i$-th component of $\mathcal{S}_{p}$, 
whereas the rest of the row is zero.
Measurements and linear coefficients for the selected sensors are denoted as $\mathbf{y}{(\mathcal{S}_{p})}=\mathbf{H}{(\mathcal{S}_{p})}\mathbf{x}$ and $\mathbf{C}{(\mathcal{S}_{p})}=\mathbf{H}{(\mathcal{S}_{p})}\mathbf{U}$, respectively.
In addition, the covariance matrix for measurement noise is expressed by $\mathbf{R}({\mathcal{S}_{p}})=\mathbf{H}{(\mathcal{S}_{p})}\mathcal{R}\mathbf{H}{(\mathcal{S}_{p})}^{\top}$.}
The argument ${(\mathcal{S}_{p})}$ will be denoted as subscript $\circ_{\mathcal{S}_{p}}$ for brevity hereinafter.

\subsection{Data-driven modeling}
\label{section: modeling}
In our implementation, the matrices $\mathbf{U}$ and $\mathcal{R}$ are generated by modal decomposition of the collected data matrix, in a process known as principal component analysis, or proper orthogonal decomposition ~\cite{brunton2019data, manohar2018data}.
The collected data $\mathbf{X} = \left[ \mathbf{x}_{1}, \,\hdots\,, \mathbf{x}_{m} \right]$ are assumed to consist of $n$-point measurements in rows by $m$ instances in columns, where $n\gg m$. 
{$\mathbf{X}$ is decomposed by singular value decomposition into $m$ orthonormal spatial modes $\mathbf{U}_{X}$ and temporal modes $\mathbf{V}_{X}$, and a diagonal matrix of singular values $\mathbf{\Sigma}_{X}$.
The approximation mode number $r$ is chosen to retain the covariance matrix for the original data matrix at a high rate:
\begin{align}
    \mathbf{X}&= \mathbf{U}_{X}\mathbf{\Sigma}_{X}\mathbf{V}_{X}^{\top}
    = \mathbf{U}\mathbf{\Sigma}\mathbf{V}^{\top}+\mathbf{U}_{N}\mathbf{\Sigma}_{N}\mathbf{V}_{N}^{\top}
    \label{eq: rom}
    , 
\end{align}
where $\mathbf{X}, \mathbf{U}_{X} \in \mathbb{R}^{n{\times}m}$ and $\mathbf{\Sigma}_{X}, \mathbf{V}_{X} \in \mathbb{R}^{m{\times}m} $, and $\mathbf{U} \in \mathbb{R}^{n{\times}r}$, $\mathbf{\Sigma} \in \mathbb{R}^{r{\times}r}$ and $\mathbf{V} \in \mathbb{R}^{m{\times}r}$, respectively.
Here, the second term with the subscript $\circ_{N}$ on the right-hand side is the portion that is excluded from $r$ rank representation and thus is regarded as the measurement noise.}
{The $i$-th column of $\mathbf{\Sigma}\mathbf{V}^{\top}$ and $\mathbf{H}_{\mathcal{S}_{p}}\mathbf{X}$ are the variable vector and the measurement in \cref{eq: measurement} for the $i$-th instance, respectively.}
Thus, one can immediately recover the low rank representation of the large-scale measurement as $\tilde{\mathbf{x}}=\mathbf{U}\tilde{\mathbf{z}}$ by obtaining an estimation $\tilde{\mathbf{z}}$~\cite{manohar2018data}.
With respect to $\mathcal{R}$, several approaches are capable of expressing the statistical behavior of the residual between the measurement and the reduced order model, $\mathbf{X}-\mathbf{U}\mathbf{\Sigma}\mathbf{V}^{\top}$, which are exemplified by kernel functions used in signal processing or data-driven modeling in Ref.~ \cite[Section 2]{yamada2021fast}.
By taking the expectation 
$\mathcal{R}=\mathbf{E}\left[\mathbf{w}\mathbf{w}^{\top}\right]$, 
the model of noise in the latter manner is denoted as $\mathcal{R}=\mathbf{U}_{N}\mathbf{\Sigma}_{N}^2 \mathbf{V}_{N}^{\top}$ of \cref{eq: rom}, which is used in \cref{section: result}.
{\Cref{append: noise} shows how the data-driven design of the measurement noise covariance is affected in the standpoint of the correlation and amount of training data.}

\subsection{Objective function for sensor selection}
Several criteria for sensor selection are available for scalar evaluation of the measurement design, like D-, E- or A-optimality mentioned in~\cite{udwadia1994methodology}.
The performances for D-, E- and A-optimality criteria were previously compared for sensor sets obtained by greedy sensor selection methods suited for these criteria~\cite{nakai2021effect}.
Sensor selection based on the D-optimality criterion performed well in both of the computational time and the values of other criteria, thus being adopted in the present study.
Note that the efficient implementation shown in \cref{section: greedy algorithm} can easily be extended to the A-optimality settings.

Geometrically, this optimization corresponds to the minimization of the volume of an ellipsoid, which represents the expected estimation error variance~\cite{joshi2009sensor}:
\begin{align}
    \argmin_{\mathcal{S}_{p}} \det\left(E\left[\left(\mathbf{z}-\tilde{\mathbf{z}}\right)
    \left(\mathbf{z}-\tilde{\mathbf{z}}\right)^{\top}\right]\right)
    \label{eq:error cov normal},
\end{align}
where the operator $E\left[ \circ \right]$ means taking the expected value of the argument.
Furthermore, this matrix is known to correspond to the inverse of the Fisher information matrix.
This equality is easily confirmed under the assumption of Gaussian measurement noise, by differentiating by $\mathbf{z}$ a log likelihood, $L=-(\mathbf{y}-\mathbf{C}_{\mathcal{S}_{p}}\tilde{\mathbf{z}})\mathbf{R}^{-1}(\mathbf{y}-\mathbf{C}_{\mathcal{S}_{p}}\tilde{\mathbf{z}})+\textrm{const.}$, then substituting estimation given by~ Eq.(\ref{eq: estimate noise over}).
The optimization returns the set of measurement point $\mathcal{S}_{p}$ from all the possible locations, although this is normally an intractable process. 
Instead, a greedy algorithm is employed with objective functions for both $p\le r$ and $p > r$.
They are derived by generalizing the formulation in Ref.~\cite{nakai2021effect} for the correlated measurement noise hereafter.
The set of sensors are evaluated only in the observable subspace of $\mathbf{R}_{\mathcal{S}_{p}}^{-1/2}\mathbf{C}_{\mathcal{S}_{p}}$, since the measurement system \cref{eq: measurement} is underdetermined when $p < r$. 
From \cref{eq:error cov normal}, the subspace is separated by the projection $\mathbf{z} \rightarrow \mathbf{\xi}=\hat{\mathbf{V}}_{\mathbf{C}}^{\top}\mathbf{z}$ after singular value decomposition of $\mathbf{R}_{\mathcal{S}_{p}}^{-1/2}\mathbf{C}_{\mathcal{S}_{p}}$:
\begin{align}
    &{E}\left[\left(\mathbf{\xi}-\tilde{\mathbf{\xi}}\right)
        \left(\mathbf{\xi}-\tilde{\mathbf{\xi}}\right)^{\top}\right] \notag\\
    &=\left\{\begin{array}{ll}
            \sigma_{\textrm{n}}^2\hat{\mathbf{U}}_{\mathbf{C}}^{\top} \left(  \mathbf{R}_{\mathcal{S}_{p}}^{-1/2}\mathbf{C}_{\mathcal{S}_{p}}\mathbf{C}_{\mathcal{S}_{p}}^{\top} \mathbf{R}_{\mathcal{S}_{p}}^{-1/2} \right)^{-1} \hat{\mathbf{U}}_{\mathbf{C}} & \left( p \le r \right) \\
            \sigma_{\textrm{n}}^2\hat{\mathbf{V}}_{\mathbf{C}}^{\top} \left(\mathbf{C}_{\mathcal{S}_{p}}^{\top}\mathbf{R}_{\mathcal{S}_{p}}^{-1}\mathbf{C}_{\mathcal{S}_{p}}\right)^{-1}\hat{\mathbf{V}}_{\mathbf{C}} & \left( p > r \right)
            \\
    \end{array}\right. \label{eq: error covarz proj}
\end{align}    
with some matrices of appropriate dimensions,
\begin{align}
    \quad \mathbf{R}_{\mathcal{S}_{p}}^{-1/2}\mathbf{C}_{\mathcal{S}_{p}}=
    \left\{\begin{array}{cl}
    \hat{\mathbf{U}}_{\mathbf{C}}
    \left[\begin{array}{cc}\hat{\mathbf{\Sigma}}_{\mathbf{C}} & \mathbf{0} \end{array}\right]
    \left[\begin{array}{cc}\hat{\mathbf{V}}_{\mathbf{C}}^{1\top} \\ \hat{\mathbf{V}}_{\mathbf{C}}^{2\top} \end{array}\right] & \left( p \le r \right)\\
    \left[\begin{array}{cc}\hat{\mathbf{U}}_{\mathbf{C}}^{1} & \hat{\mathbf{U}}_{\mathbf{C}}^{2} \end{array}\right]
    \left[\begin{array}{c}\hat{\mathbf{\Sigma}}_{\mathbf{C}}\\\mathbf{0}\end{array}\right] \hat{\mathbf{V}}_{\mathbf{C}}^{\top} & \left( p > r \right).
    \end{array}\right. \notag
\end{align}
{The evaluation of the error covariance in the observable subspace was recently introduced by Nakai et al.~\cite{nakai2021effect}, and it is extended to the correlated noise case in the present study for the first time (to the best of our knowledge). }
One can use various metrics for the projected covariance matrix \cref{eq: error covarz proj} like its determinant, trace or minimum eigenvalue, as in Ref.~\cite{joshi2009sensor, nakai2021effect}. 
The determinant of the inverse matrices in \cref{eq: error covarz proj} is maximized in the present manuscript:
\begin{subnumcases}{\label{eq: det noise re}}
    \argmax_{\mathcal{S}_{p}}\, \det\left(\mathbf{R}_{\mathcal{S}_{p}}^{-1}\mathbf{C}_{\mathcal{S}_{p}}\mathbf{C}_{\mathcal{S}_{p}}^{\top}\right) &$\left( p\le r \right) \label{eq: det noise under} $\\
    \argmax_{\mathcal{S}_{p}}\, \det\left(\mathbf{C}_{\mathcal{S}_{p}}^{\top}\mathbf{R}_{\mathcal{S}_{p}}^{-1}\mathbf{C}_{\mathcal{S}_{p}}\right) &$\left( p > r \right) \label{eq: det noise over}$.
\end{subnumcases}
{Note that whitening all candidates with $\mathcal{R}^{-1}$ before sensor selection is based on the assumption of weakly correlated noise, because $\mathcal{R}$ contains noise covariance over sensors that are not selected as pointed out in \cite{liu2016sensor}.
In \cref{section: greedy algorithm}, an algorithm is presented for achieving \cref{eq: det noise over} in a greedy manner.}

\subsection{Efficient greedy algorithm} \label{section: greedy algorithm} 
\Cref{alg: correlated} shows the procedure implemented in the computation conducted in \cref{section: result}, which implicitly exploits the one-rank determinant lemma as \cite{liu2016sensor, saito2021determinant, yamada2021fast}. It is worth mentioning that the previously presented noise-ignoring algorithm in \cite{saito2021determinant}, \cref{alg: uncorrelated}, is easily obtained by substituting an identity matrix into $\mathcal{R}$.
\begin{algorithm}
\caption{Determinant-based greedy algorithm with noise covariance matrix {(DG/NC)}}\label{alg: correlated}
\begin{algorithmic}
\renewcommand{\algorithmicrequire}{\textbf{Input:}}
\renewcommand{\algorithmicensure}{\textbf{Output:}}
    \REQUIRE $\mathbf{U}\in\mathbb{R}^{n \times r},\, \mathcal{R}\in\mathbb{R}^{n \times n},\, p>0$
    \ENSURE Indices of chosen $p$ sensor positions $\mathcal{S}_{p}$
    \STATE 
    $
    \mathcal{S}_{n} \leftarrow \left\{ 1, \hdots, n\right\},\,
    \mathcal{S}_{0} \leftarrow \emptyset
    $
    \FOR{ $k =1, \dots, p$ }
    \IF{$k \leq r$}
          \STATE $ i_{k} = \underset{i\, \in\, \mathcal{S}_{n}\, \backslash\, \mathcal{S}_{k-1}}{\argmax} \det(\mathbf{R}_{\mathcal{S}_{k-1}\cup\, i}^{-1}\mathbf{C}_{\mathcal{S}_{k-1}\cup\, i} \mathbf{C}_{\mathcal{S}_{k-1}\cup\, i}^{\top})$ ... [\cref{eq: 1rank under}]\\
    \ELSE
          \STATE $ i_{k} = \underset{i\, \in\, \mathcal{S}_{n}\, \backslash\, \mathcal{S}_{k-1}}{\argmax} \det(\mathbf{C}_{\mathcal{S}_{k-1}\cup\, i}^{\top} \mathbf{R}_{\mathcal{S}_{k-1}\cup\, i}^{-1} \mathbf{C}_{\mathcal{S}_{k-1}\cup\, i})$ ... [\cref{eq: 1rank over}]\\
    \ENDIF
    \STATE 
    $\mathcal{S}_{k} \leftarrow \mathcal{S}_{k-1} \cup i_{k}$
    \ENDFOR
\end{algorithmic}
\end{algorithm}

\begin{algorithm}
\caption{Determinant-based greedy algorithm {(DG)\cite{saito2021determinant}}\label{alg: uncorrelated}}
\begin{algorithmic}
\renewcommand{\algorithmicrequire}{\textbf{Input:}}
\renewcommand{\algorithmicensure}{\textbf{Output:}}
    \REQUIRE $\mathbf{U}\in\mathbb{R}^{n \times r},\, p>0$
    \ENSURE Indices of chosen $p$ sensor positions $\mathcal{S}_{p}$
    \STATE 
    $
    \mathcal{S}_{n} \leftarrow \left\{ 1, \hdots, n\right\},\,
    \mathcal{S}_{0} \leftarrow \emptyset
    $
    \FOR{ $k =1, \dots, p$ }
    \IF{$k \leq r$}
          \STATE $ i_{k} = \underset{i\, \in\, \mathcal{S}_{n}\, \backslash\, \mathcal{S}_{k-1}}{\argmax} \det(\mathbf{C}_{\mathcal{S}_{k-1}\cup\, i} \mathbf{C}_{\mathcal{S}_{k-1}\cup\, i}^{\top})$\\
    \ELSE
          \STATE $ i_{k} = \underset{i\, \in\, \mathcal{S}_{n}\, \backslash\, \mathcal{S}_{k-1}}{\argmax} \det(\mathbf{C}_{\mathcal{S}_{k-1}\cup\, i}^{\top} \mathbf{C}_{\mathcal{S}_{k-1}\cup\, i})$\\
    \ENDIF
    \STATE 
    $\mathcal{S}_{k} \leftarrow \mathcal{S}_{k-1} \cup i_{k}$
    \ENDFOR
\end{algorithmic}
\end{algorithm}
The equations are converted by the lemma shown later herein. First, consider an objective function when there are fewer sensors deployed than the number of state variables $(p \le r)$:
\begin{align}
    &\det\left(\mathbf{R}_{\mathcal{S}_{k-1}\cup\, i}^{-1}\mathbf{C}_{\mathcal{S}_{k-1}\cup\, i}\mathbf{C}_{\mathcal{S}_{k-1}\cup\, i}^{\top}\right) \notag\\
    &= \det\left(\mathbf{R}_{\mathcal{S}_{k-1}\cup\, i}^{-1}\right)\, \det\left(\mathbf{C}_{\mathcal{S}_{k-1}\cup\, i}\mathbf{C}_{\mathcal{S}_{k-1}\cup\, i}^{\top}\right) \notag\\
    &= \frac{
    \begin{aligned}
    \left(\mathbf{u}_{i}\mathbf{u}_{i}^\text{T}-\mathbf{u}_{i}\mathbf{C}_{\mathcal{S}_{k-1}}^\text{T} \left(\mathbf{C}_{\mathcal{S}_{k-1}}\mathbf{C}_{\mathcal{S}_{k-1}}^\text{T}\right)^{-1} \mathbf{C}_{\mathcal{S}_{k-1}}\mathbf{u}_{i}^\text{T}\right)\\
    \end{aligned}
    }{
    \left(\mathbf{t}_{i}-\mathbf{s}_{k(i)} \mathbf{R}_{\mathcal{S}_{k-1}}^{-1}\mathbf{s}_{k(i)}^{\rm{T}}\right)\det\left[{\left(\mathbf{R}_{\mathcal{S}_{k-1}}^{-1}\mathbf{C}_{\mathcal{S}_{k-1}}\mathbf{C}_{\mathcal{S}_{k-1}}^\text{T}\right)}^{-1}\right]
    } \label{eq: greedy under}, 
\end{align}
where the subscript ${k(i)}$ represents the component produced by the $i$-th sensor candidate in the $k$-th step:
\begin{align}
    &\begin{cases}
    \mathbf{C}_{\mathcal{S}_{k-1}\cup\, i} = \left(\begin{array}{c}\mathbf{C}_{\mathcal{S}_{k-1}}\\\mathbf{u}_{i}\end{array}\right),\\
    \mathbf{R}_{\mathcal{S}_{k-1}\cup\, i} = \left(\begin{array}{cc}\mathbf{R}_{\mathcal{S}_{k-1}}&\mathbf{s}_{k(i)}^{\top}\\\mathbf{s}_{k(i)}&{t}_{i}\end{array}\right).
    \end{cases}
    \notag
\end{align}
Here, $\mathbf{u}_{i}$ is the $i$-th row of $\mathbf{U}$, and $\mathbf{s}_{k(i)}$ and ${t}_{i}$ are the noise covariance between the selected sensors given by the previous steps and the $i$-th candidate and noise variance for the $i$-th candidate, respectively.
The algorithm avoids expensive computations involving the determinant by separating the components of the obtained sensors from the objective function in the current selection step of \cref{eq: greedy under}:
\begin{align}
    &i_{k}=  \underset{i\, \in\, \mathcal{S}_{n}\, \backslash\, \mathcal{S}_{k-1}}{\argmax}\,\det\left(\mathbf{R}_{\mathcal{S}_{k-1}\cup\, i}^{-1}\right)\det\left(\mathbf{C}_{\mathcal{S}_{k-1}\cup\, i}\mathbf{C}_{\mathcal{S}_{k-1}\cup\, i}^{\top}\right)  \notag\\
    &=\underset{i\, \in\, \mathcal{S}_{n}\, \backslash\, \mathcal{S}_{k-1}}{\argmax}\,
    \frac{\mathbf{u}_{i}\left(\mathbf{I}-\mathbf{C}_{\mathcal{S}_{k-1}}^\text{T}\left(\mathbf{C}_{\mathcal{S}_{k-1}}\mathbf{C}_{\mathcal{S}_{k-1}}^\text{T}\right)^{-1}\mathbf{C}_{\mathcal{S}_{k-1}}\right)\mathbf{u}_{i}^\text{T}
    }{
    \mathbf{t}_{i}-\mathbf{s}_{k(i)} \mathbf{R}_{\mathcal{S}_{k-1}}^{-1}\mathbf{s}_{k(i)}^{\rm{T}}},\label{eq: 1rank under} 
\end{align}
and then a unit vector $\mathbf{e}_{i_{k}} \in \mathbb{R}^{1{\times}n}$, of which only the $i_{k}$-th entry is unity, is added to the $k$-th row of $\mathbf{H}$.
Note that \cref{eq: 1rank under} corresponds to maximization of the difference when an arbitrary sensor is added to the sensor set of the previous step. The numerator of \cref{eq: 1rank under} is the $\ell_2$ norm of the vector, and the denominator is positive, because the covariance matrix $\mathbf{R}_{\mathcal{S}_{k-1}}$ is assumed to be positive definite. 
Subsequently, the objective function is modified for the case in which more sensors than the number of state variables have already been determined:
\begin{align}
&i_{k}= \underset{i\, \in\, \mathcal{S}_{n}\, \backslash\, \mathcal{S}_{k-1}}{\argmax} \, \det\left(\mathbf{C}_{\mathcal{S}_{k-1}\cup\, i}^{\top}\mathbf{R}_{\mathcal{S}_{k-1}\cup\, i}^{-1}\mathbf{C}_{\mathcal{S}_{k-1}\cup\, i}\right) \notag 
\\
&=\underset{i\, \in\, \mathcal{S}_{n}\, \backslash\, \mathcal{S}_{k-1}}{\argmax}\,{
     \frac{
        \mathbf{\phi}_{(i)} \left(\mathbf{C}_{\mathcal{S}_{k-1}}^{\top}\mathbf{R}_{\mathcal{S}_{k-1}}^{-1}\mathbf{C}_{\mathcal{S}_{k-1}}\right)^{-1} \mathbf{\phi}_{(i)}^{\top}
        }{
        {t}_{i}-\mathbf{s}_{k(i)}\mathbf{R}_{\mathcal{S}_{k-1}}^{-1}\mathbf{s}_{k(i)}^{\top}
        },} \label{eq: 1rank over}
\end{align}
where $
\mathbf{\phi}_{(i)}=\mathbf{s}_{k(i)}\mathbf{R}_{\mathcal{S}_{k-1}}^{-1}\mathbf{C}_{\mathcal{S}_{k-1}}-\mathbf{u}_{i}. 
$
\cref{eq: 1rank over} is 
positive, and the objective function, Eq.~\eqref{eq: det noise over}, increases monotonically. Details of the expansion are found in Ref.~\cite{yamada2021fast}.

The computational cost of algorithms are listed in \cref{tab: methods}:
\begin{table}[!hptb]
    \centering
    \subfloat[][Methods for uncorrelated noise]{
    \label{tab: methods previous}
    \begin{tabular}{p{150pt} p{120pt}}
        Algorithm & Complexity
        \\ \hline
        Greedy algorithm~\cite{saito2021determinant} & $\mathcal{O}(pnr^{2})$
        \\
        Alternating direction& \multirow{2}{*}{$\mathcal{O}(nr^{2})$ / iteration}
        \\ \multicolumn{1}{r}{method of multipliers~\cite{nagata2021data}} & 
        \\
        Newton method~\cite{joshi2009sensor}&$\mathcal{O}(n^{3})$ / iteration
        \\
        Newton method with &  \multirow{2}{*}{$\mathcal{O}(\tilde{n}^{3})$ / iteration\quad $\left( \tilde{n}<n\right)$}
        \\\multicolumn{1}{r}{randomized subspace~\cite{nonomura2021randomized}}& 
        \\ \hline
    \end{tabular}}
    \\
    \subfloat[][Methods for correlated noise]{
    \label{tab: methods proposed}
    \begin{tabular}{p{150pt} p{120pt}}
        Algorithm & Complexity 
        \\ \hline
        Alternating direction& \multirow{2}{*}{$\mathcal{O}(nm^{2})$ / iteration}
        \\ \multicolumn{1}{r}{method of multipliers~\cite{nagata2022data}} & 
        \\
        Semi definite programming~\cite{liu2016sensor}&  $\mathcal{O}(n^{4.5})$ / iteration
        \\
        Greedy algorithm (\bf{Proposed})& $\mathcal{O}(pnm^{2})$
        \\ \hline
    \end{tabular}}
    \captionsetup{format=plain,font=normal,margin=10pt,name=Table}
    \caption{\reviewerA{Sensor selection algorithms using various optimization metrics}
    \label{tab: methods}}
\end{table}

The objective function loses submodularity if the measurement noises at different sensor positions are strongly correlated with each other (or when the off-diagonal components in $\mathbf{R}$ are no smaller than the diagonal components.)
The following example provides a nonsubmodular and nonsupermodular case for Eq.~\eqref{eq: det noise over}.
For simplicity, the spatial modes $\mathbf{U} \in \mathbb{R}^{3}$ and the noise covariance $\mathcal{R} \in \mathbb{R}^{3{\times}3}$ are set as follows. Here, the noise components $i=2,3$ are strongly correlated, whereas those for $i=1$ are relatively independent.
\begin{align}
    \mathbf{U} &= \left( \begin{array}{c}
        0.1 \\
        1 \\
        1
    \end{array} \right), \quad
    \mathcal{R} = \left( \begin{array}{ccc}
        1   & -0.1 & 0.1 \\
        -0.1 & 0.8 & 0.7 \\
        0.1 & 0.7 & 2 
    \end{array} \right) \notag
\end{align}
With these matrices, the values of the determinant function Eq.~\eqref{eq: det noise over} are
\begin{align}
    f_{\{ 1,2   \}} - f_{\{ 1     \}} = 1.2913 &> f_{\{ 1,2,3 \}} - f_{\{ 1,3   \}}=0.8038 \nonumber\\ 
    f_{\{ 1,3   \}} - f_{\{ 3     \}} = 0.0025 &< f_{\{ 1,2,3 \}} - f_{\{ 2,3   \}}=0.0450,\nonumber
\end{align}
where $f_{\mathcal{S}}\, (\mathcal{S} \in {2}^{\{1,2,3\}})$ refers to the value of the determinant Eq.~\eqref{eq: det noise over} for the power sets of selected sensors.
This example immediately shows that the objective function 
Eq.~\eqref{eq: det noise over} has neither submodularity nor supermodularity, whereas submodularity exists for the case with equally distributed uncorrelated measurement noise \cite{saito2021determinant}.
Thus, the sensor selection problem Eq.~\eqref{eq: det noise re} with a greedy method generally has no performance guarantee based on submodularity or supermodularity.

\section{Results}\label{section: result}
This section describes some experiments that validate the algorithm. First, data matrices are constructed from randomly generated orthonormal bases. The NOAA-SST dataset~\cite{noaa} shows the results of a practical application. 

\begin{figure}[!t]
\centering
\includegraphics[width=0.6\linewidth]
{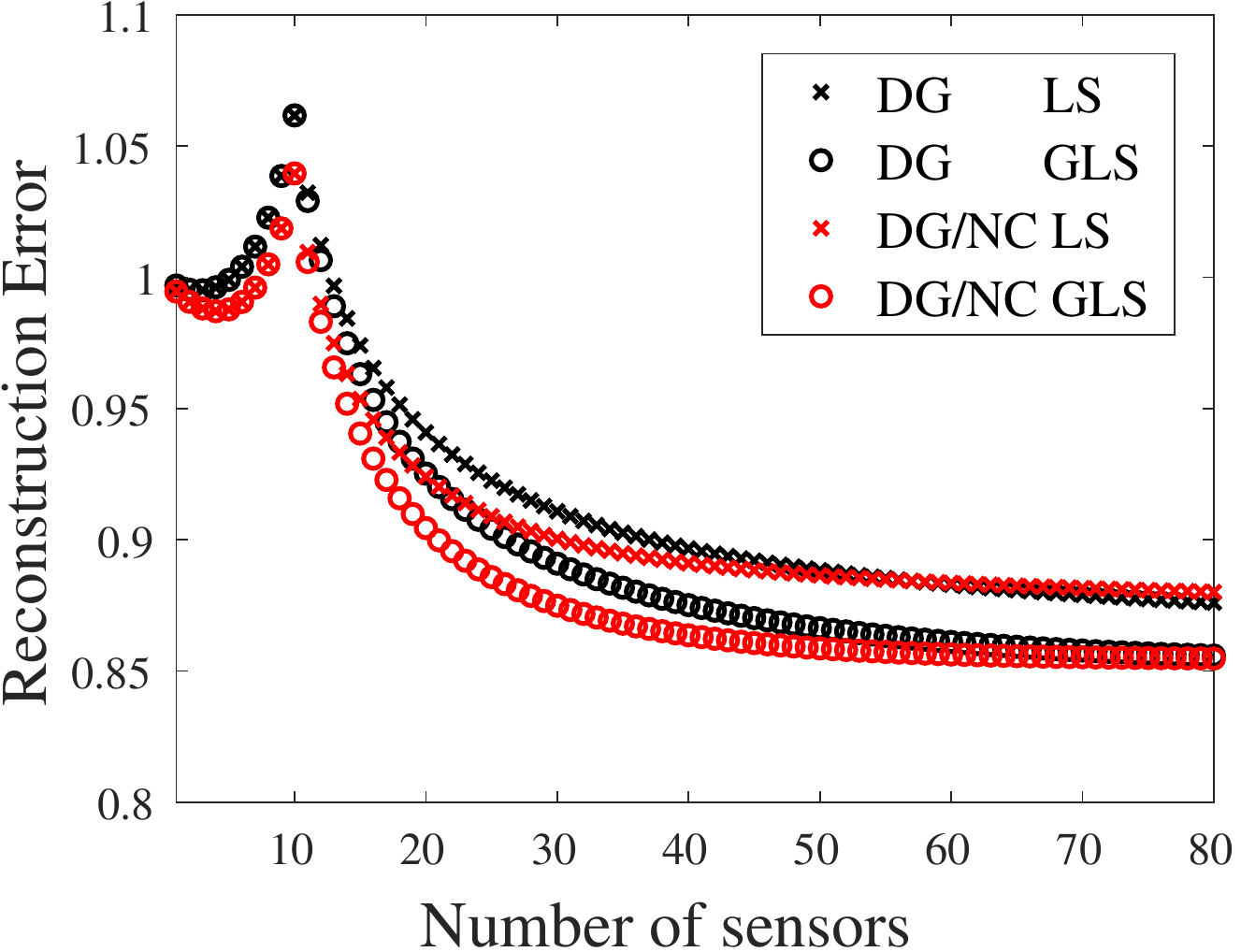}
\caption{Estimation error comparison of two selection algorithms and two estimation methods, averaging 2,000 tests. (Labels: DG and DG/NC are the previously presented and the proposed algorithm for sensor selection, LS and GLS are normal linear least squares estimation and generalized estimation by considering noise correlation, respectively)}
\label{fig:error_random}
\end{figure}

\subsection{Randomly generated data matrix} \label{section: random}
Generalized results are shown in this subsection.
The problem considered here is as follows: a data matrix $\mathbf{X}$ is constructed as $\mathbf{X}=\mathbf{U}_{X}\Sigma_{X}\mathbf{V}_{X}^{\top}$, where $\mathbf{U}_{X}$ and $\mathbf{V}_{X}$ are $5,000 \times 100$ and $100 \times 100$ orthonormal matrices, respectively, generated from appropriately sized matrices containing numbers from a standard normal distribution, and $\Sigma_{X}$ is a diagonal matrix with $\diag\left(\Sigma_{X}\right) = \left(\begin{array}{llllll}
 1 & 0.99 & \hdots & (101-j)/100 & \hdots & 0.01
\end{array}\right)
$.
The algorithms for the sensor selection treat these matrices after dividing the first 10 columns as $\mathbf{U}$ and $\mathbf{V}$ and the remaining columns as $\mathbf{U}_{N}$ and $\mathbf{V}_{N}$. Then, the first 10 diagonal components and the remaining are labeled as $\mathbf{\Sigma}$  and $\mathbf{\Sigma}_{N}$, respectively.
Note that the measure $e$ in terms of ``reconstruction error'' is expressed as 
$    e = {\|\mathbf{X}-\mathbf{U}\tilde{\mathbf{Z}}\|_{\mathrm{F}}} / {\|\mathbf{X}\|_{\mathrm{F}}}. \notag $
Here, the series for the estimation $\tilde{\mathbf{z}}$ of \cref{eq: estimation gls} is concatenated as $\tilde{\mathbf{Z}}$, and $\|\circ\|_{\mathrm{F}}$ represents the Frobenius norm of $\circ$.
\Cref{fig:error_random} shows the result of the reconstruction with the estimate with $p$ sensors and the $r$ dimensional reduced-order model \cref{eq: rom}.
Here, DG and DG/NC in the legend refer to ``determinant-based greedy algorithm'' in Ref.~\cite{saito2021determinant} and \cref{alg: correlated} considering ``noise covariance'' in the measurement, respectively, and LS and GLS refer to ``linear least squares estimation'' and ``generalized linear least squares estimation'' using noise covariance, respectively.
Note that the plots for $p \le r$ are calculated by the same estimator Eq.~(\ref{eq: estimate noise under}), and, therefore, the estimations with a small number of sensors for both GLS and LS are identical for each selected sensor set.
First, the GLS estimation reduces the reconstruction error in oversampling cases for sensors for both algorithms.
The measurement noise is quite excessive, and thus, sensors of both selection methods exhibit comparable results for the LS estimation.
Second, the more sensors are deployed, the lower the reduction becomes thanks to the GLS estimation.
This is partly because measurement using a large number of sensors suppresses outliers resulting from the correlated measurement noise.
If a much larger number of sensors is available than the number of estimated variables, the importance of correlation in the measurement noise might diminish.
\subsection{NOAA-SST} \label{section: noaasst}
\begin{figure}[!t]
\centering
    \subfloat[Previous greedy algorithm w/ $\mathbf{U}$] 
    {\label{fig: sensors sst_dgwr} \includegraphics[clip,width=0.6\linewidth]{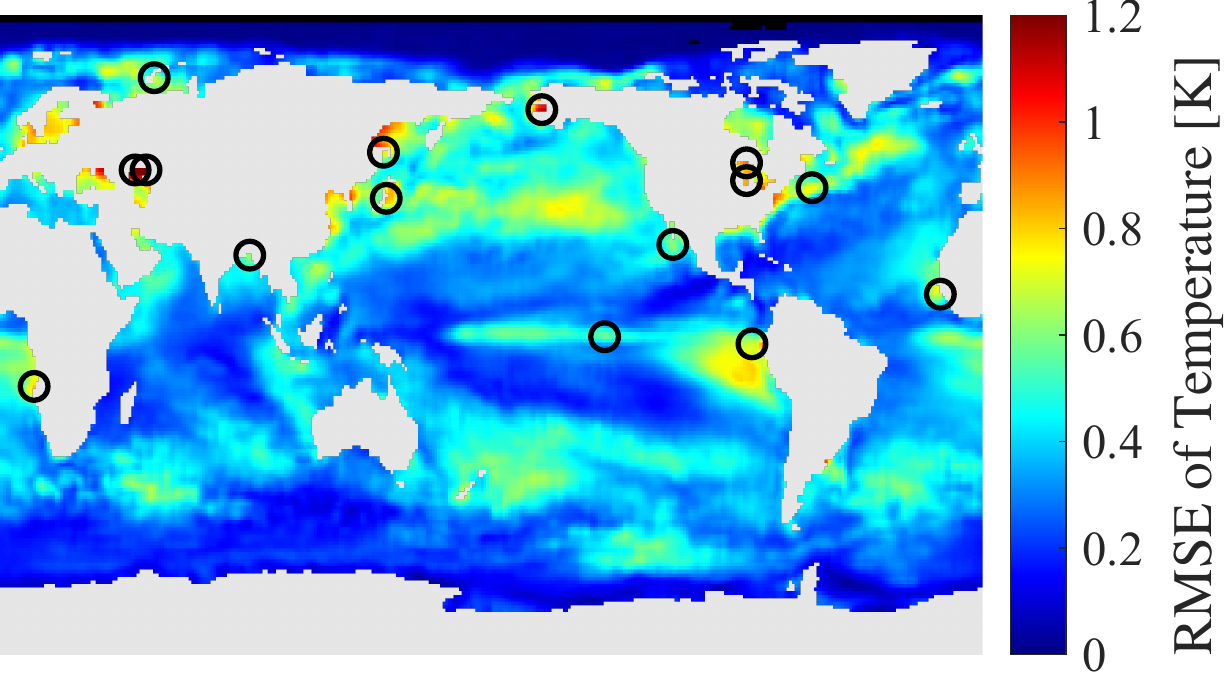}}\\
    \subfloat[Presented greedy algorithm  w/ $\mathbf{U}$ and $\mathcal{R}$] {\label{fig: sensors sst_dg} \includegraphics[clip,width=0.6\linewidth]{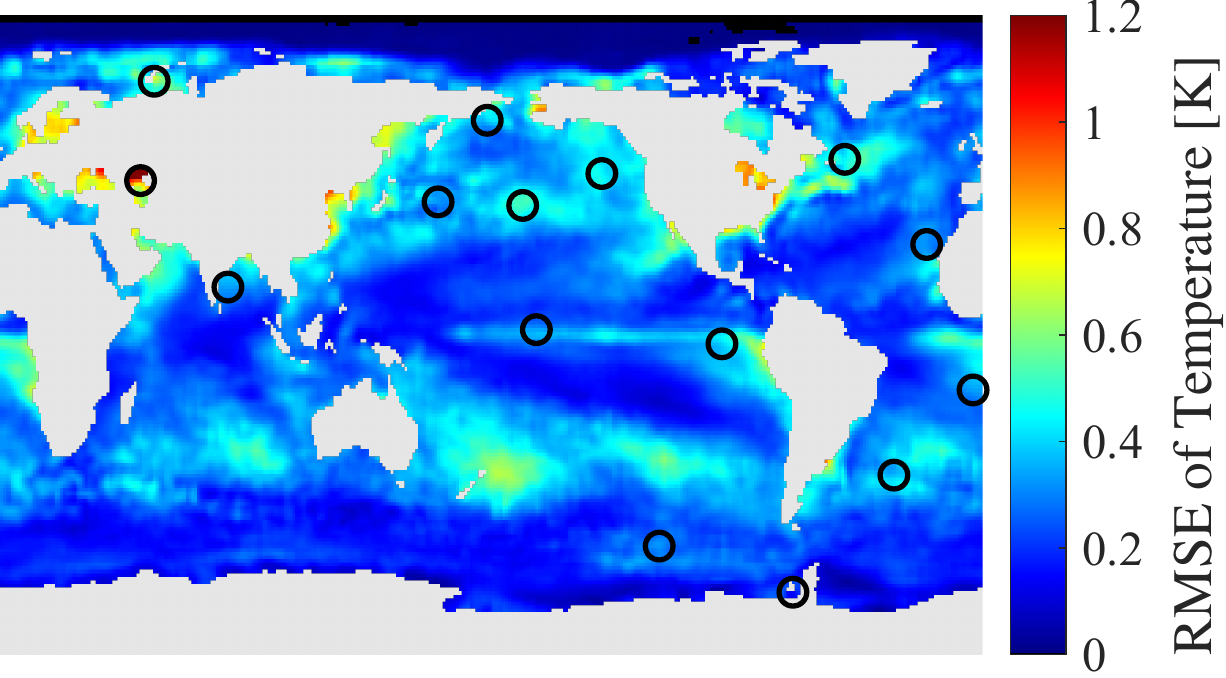}}
    \caption{Fifteen sensors for global distribution of sea surface temperature selected by the greedy algorithms, on color maps of the root mean square of the estimation error $\mathbf{U}\mathbf{\Sigma}\mathbf{V}^{\top} - \mathbf{U} \tilde{\mathbf{z}}$ using each sensor set. The sensors of the presented algorithm are distributed and show better performance.}
\label{fig:sensors sst}
\end{figure}

Here, we apply this strategy to pursue sensor selection using large-dimensional climate data. A brief description of the NOAA-SST data is given in \cref{table:SST_data_conditions}.

\begin{table}[!htbp]
\centering
\caption{Description of SST data.} 
\label{table:SST_data_conditions}
\begin{tabular}{p{100pt}p{300pt}}
\hline
Label & NOAA Optimum Interpolation (OI) SST V2 \cite{noaa}
\\
Temporal Coverage
& {Weekly means from 1989/12/31 to 1999/12/12 ($520$ snapshots)}\\
Spatial Coverage& 
{1.0 degree latitude $\times$ 1.0 degree longitude
global grid ($n=44219$ measurement on the ocean)}\\
\hline
\end{tabular}
\end{table}
Similar to \cref{section: random}, orthonormal modes are prepared by conducting SVD on the data matrix with the average being subtracted, then the first 10 of the 520 modes are used to build the reduced-order model of the temperature distribution ($r=10$).
The remaining modes are used for the noise covariance matrix.
Several sensor positions for which the noise amplitude is extremely low (smaller than 1\% of the maximum RMS of noise in this comparison) are eliminated from candidate set $\mathcal{S}_{n}$ beforehand, as conducted in \cite{yamada2021fast}. 
The results in this subsection can be compared with those in Ref.~\cite{manohar2018data}, \cite{yamada2021fast}, or \cite{saito2021determinant}.
In \cref{fig:sensors sst}, the positions of sensors are represented by open circles on the colored maps, which illustrate the fluctuation of estimation error using those sensors, namely $\mathbf{U}\mathbf{\Sigma}\mathbf{V}^{\top} - \mathbf{U} \tilde{\mathbf{z}} $.
The difference in the sensor positions is remarkable, since the proposed algorithm spreads sensors and avoids neighboring sensors that might be affected by correlated measurement noise.
The reduction in the estimation error is also recognizable by comparing the backgrounds.

\begin{figure}[!t]
\centering
 \includegraphics[width=0.6\linewidth]{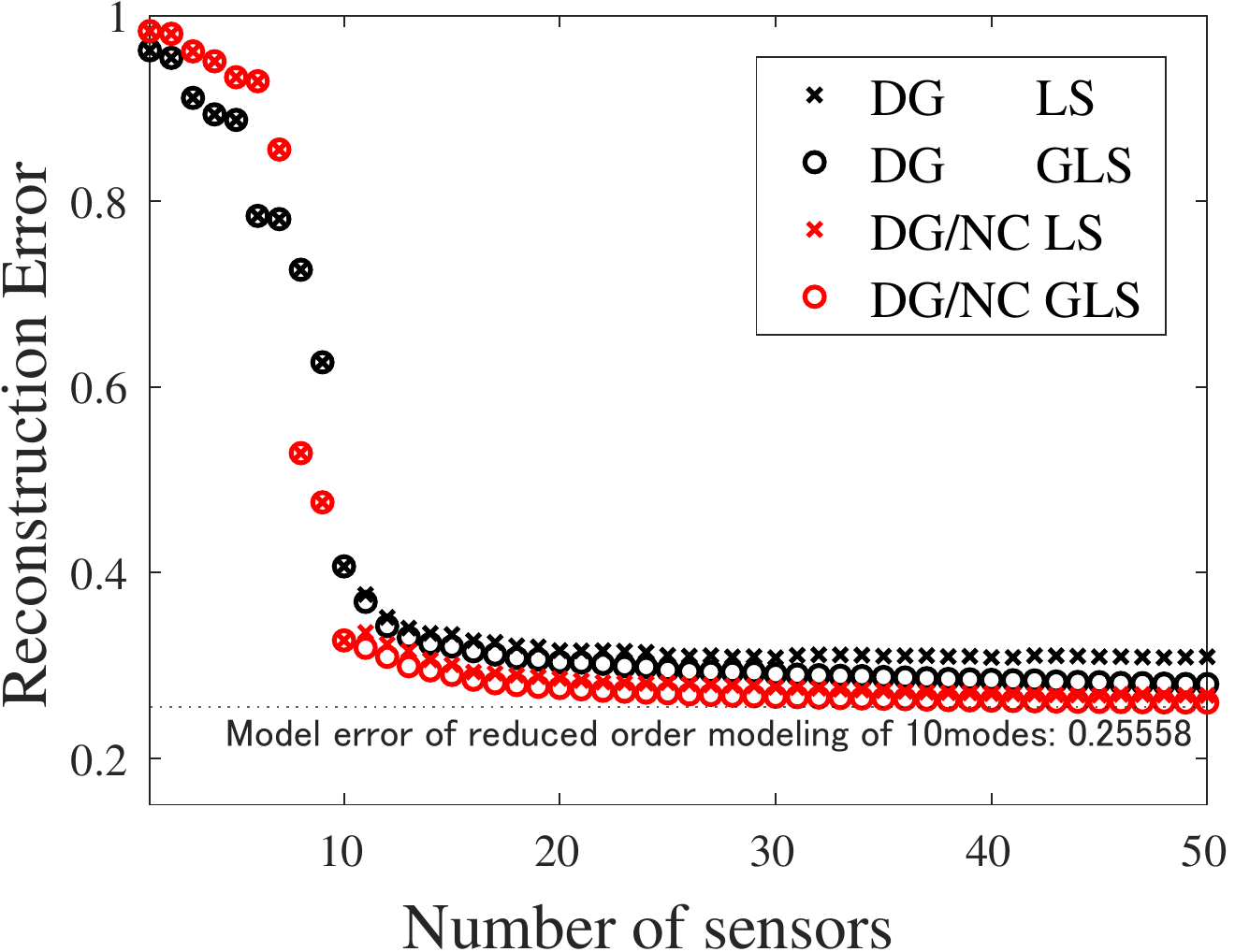}
\caption{Estimation error comparison of two selection algorithms and two estimation methods for NOAA-SST data:. (Labels: DG and DG/NC are the previously presented and the proposed algorithm for sensor selection, LS and GLS are normal linear least squares estimation and generalized estimation by considering noise correlation, respectively}
\label{fig:error_SST}
\end{figure}

\Cref{fig:error_SST} compares the results of estimation using the noise covariance information.
Note that cross-validation is not conducted for this comparison, since it is hard to quantify the estimation error because of the dynamics in SST which is partly extracted by the reduced order modeling.
In \cref{append: noise}, the covariance matrix of measurement noise is characterized by the number of snapshots to form the noise covariance matrix.
A horizontal broken line in \cref{fig:error_SST} shows the modeling error due to the low-rank representation of \cref{eq: rom}.
The red plots show better performance for sensors using the proposed algorithm than those using the previous DG algorithm~\cite{saito2021determinant} owing to the noise covariance matrix in the sensor selection procedure.
There are several differences in the trend of plots compared to \cref{fig:error_random}, {\it e.g.}\, the contribution from sensors of the proposed algorithm is more significant than that for the GLS estimation.
This is perhaps because of the weak amplitude of higher ordered modes in addition to the similarity in the location where the reduced-order phenomena and the measurement noise fluctuates greatly.
The proposed algorithm that involves noise covariance evaluates the positions with less measurement noise. 
Therefore, accurate estimation is achieved even with linear least squares estimation, which contrasts with the errors of sensors of DG algorithm staying relatively high.

\section{Conclusions}\label{conclusion}
A greedy algorithm for sensor selection for generalized least squares estimation is presented. A covariance matrix generated by truncated modes in reduced-order modeling is applied and a weighting matrix is built for the estimation. A specialized one-rank lemma involving the covariance matrix realizes a simple transformation from the true optimization into a series of a greedy scalar evaluation. In addition, the objective function is shown to be neither submodular nor supermodular. 
Numerical tests using two kinds of datasets are performed to assess the proposed determinant-based optimization method.
The proposed algorithm gives less noisy sensors and results in stable estimation in the presence of measurement noise from truncated modes of the reduced-order modeling.


\section*{Acknowledgements}
The present study was supported by JSPS KAKENHI (21J20671), JST ACT-X (JPMJAX20AD), JST CREST (JPMJCR1763) and JST FOREST (JPMJFR202C).
\appendices
\section{Data-driven noise correlation in real-world data}\label{append: noise}
In this section, some numerical experiments are conducted and the data dependency of the data-driven modeling of the measurement noise are explored. 
Mainly impact of the number of the used snapshots is investigated in this section.
Here, 
six-fold cross-validation is applied to 624 snapshots of the same NOAA-SST data used in \cref{section: noaasst}.

\begin{figure}[!ht]
\centering
 \includegraphics[width=0.6\linewidth]{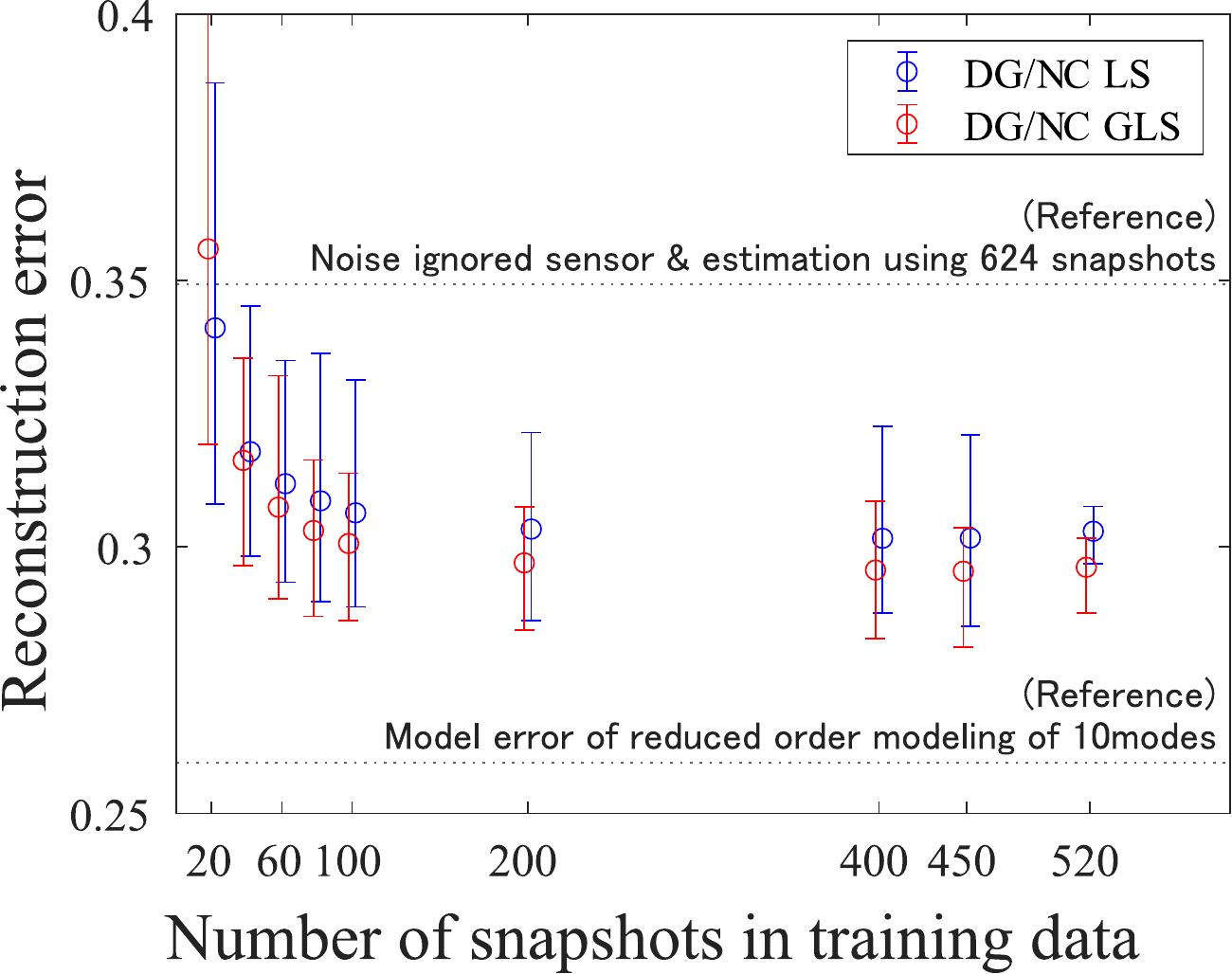}
\caption{Comparison of estimation error for NOAA-SST data using 15 sensors determined: 
least squares estimation and generalized estimation with noise considered sensors using different number of snapshots. (Circle: average of cross-validation and 50 times resampling; Error bars: maximum and minimum)}
\label{fig: error_Rtest}
\end{figure}

The procedure is summarized as follows:
\begin{enumerate}
    \item Save 624 snapshots in the memory of computer
    \item Calculate reduced order representation $\bf{U}$ by \cref{eq: rom}
    \item Randomize the order of the snapshots and divide into six parts
    \item Sample the predetermined number of snapshots randomly from 520 snapshots labeled as ``training'' then calculate $\mathcal{R}$ \label{item: calc noise}
    \item Determine 15 sensor positions using $\mathbf{U}$ and $\mathcal{R}$
    \item Reconstruct all-points measurement of 104 snapshots labeled as ``test'' using determined sensors and corresponding $\mathbf{R}$  \label{item: reconst}
    \item Store reconstruction error
    \item Resample snapshots for calculating $\mathcal{R}$, then repeat from \cref{item: calc noise} to \cref{item: reconst} for 50 times \label{item: resample}
    \item Change ``training'' and ``test'', then repeat from \cref{item: calc noise} to \cref{item: resample}
\end{enumerate}
Here, the low-rank representation is fixed for all of the sampling cases, and the change in the reduced order representation $\mathbf{U}$ which reflects temperature dynamics is excluded.  
An evaluation of the quality including the reduced order model needs more profound discussion, and thus, this topic remains to be solved.
Calculation of noise correlation matrix in \cref{item: calc noise} above is carried out by taking $\mathcal{R} := \mathbf{X}_{N}\mathbf{X}_{N}^\top$, where $\mathbf{X}_{N} = \left( \mathbf{I} - \mathbf{U}\mathbf{U}^\top \right)\mathbf{X}$ with some notations in \cref{section: sparse,section: modeling}.

In \cref{fig: error_Rtest}, the result is summarized by the average, maximum, and minimum of the reconstruction error with the abscissa of the number of snapshots used as training data. 
Among the two horizontal broken lines, the top one corresponds to the average of the reconstruction error of six divided snapshots by the previous approach that only uses $\mathbf{U}$, and the bottom to the modeling error by approximating original snapshots by $r = 10$ modes.
The lack of the training data for calculating $\mathcal{R}$ influences the reconstruction at 20 snapshots, possibly since the weighting term using measurement noise is not captured well.
The presented method, however, results in better performance than the previous approach as the number of snapshots in the training data increases.
For the further reduction in the reconstruction error, it seems effective to increase the number of sensors (as shown in \cref{section: result}), or consider the dynamics of the measured phenomena to the estimation if the time-series snapshots are applied.

\bibliography{xaerolab_1205}

\end{document}